\documentclass[12pt,letterpaper]{article}
\begin{document}

\newcommand{\be}{\begin{equation}}
\newcommand{\ee}{\end{equation}}

\author{H.~Neuberger\\ [7mm]
  {\normalsize\it Department of Physics and Astronomy, Rutgers University}\\
  {\normalsize\it Piscataway, NJ 08855, U.S.A} }

\title{
Burgers' equation in 2D SU(N) YM.}
\maketitle \vskip 1.5cm

\abstract {\noindent It is shown that the logarithmic derivative of the characteristic polynomial of a Wilson loop in two dimensional pure Yang Mills theory with 
gauge group SU(N) exactly satisfies Burgers' equation, with 
viscosity given by 1/(2N). The Wilson loop does not intersect itself and Euclidean
space-time is assumed flat and infinite. This result 
provides a precise framework in 2D YM for recent observations 
of Blaizot and Nowak and was inspired by their work.}

\bigskip
\newpage

\tableofcontents

\vskip 1cm

\section{Introduction.}

Recent numerical work provides evidence
that Wilson loops in $SU(N)$ gauge theory in two, three and four dimensions
exhibit an infinite $N$ phase transition as they are dilated from a 
small size to a large one; in the course of this dilation the eigenvalue distribution of the untraced Wilson loop unitary matrix expands from a small 
arc on the unit circle to encompassing the entire unit circle~\cite{ourjhep, three-d}. 
An analogous effect takes place in the two dimensional 
principal chiral model for $SU(N)$~\cite{pcm}.

The universality class of this transition is 
that of a random multiplicative ensemble of unitary matrices.
The transition was discovered by Durhuus and Olesen~\cite{duol} (DO) when they solved the
Makeenko-Migdal~\cite{makeenko} loop equations in two dimensional planar QCD. 
The associated multiplicative random matrix ensemble~\cite{janik} can be axiomatized in the
language of noncommutative probability~\cite{voicu}. It provides
a generalization of the familiar law of large numbers.
The essential feature making a difference is that 
one case is commutative and the other not. 
Various recent insights into the DO transition
~\cite{olesena, olesenb, blaizot} point to 
possibly deeper interpretations of the transition. 

In this note, motivated by a recent paper 
by Blaizot and Nowak~\cite{blaizot}, I present an
exact map from the average characteristic polynomial associated with
a Wilson loop to Burgers' equation. This extends to finite $N$ 
the original work of DO at $N=\infty$, where the {\sl inviscid} Burgers' equation plays a central role. The main observation is that all finite $N$ effects are exactly 
represented by reinstating a finite viscosity in Burgers' equation, given by $\frac{1}{2N}$. Positive $N$ gives positive viscosity, so the equation knows
at least that $N$ should not be negative. I suspect that integral  
$N$'s are identified as special by a
Mittag-Leffler~\cite{marku} representation 
of the solution, stemming from a product representation of the average characteristic
polynomial, and depending also on the initial condition. 

In addition to making the 
insight of~\cite{blaizot} particularly transparent, I hope that this result 
would also aid future efforts to exploit 
large $N$ universality in dimensions higher than two for obtaining analytical 
quantitative estimates of the ratio
between a scale describing perturbative
phenomena and the scale of confinement. This was the original motivation 
for seeking to establish numerically 
large $N$ phase transitions in Wilson loops~\cite{ourjhep}. 

\section{Characteristic polynomial.}

An $N \times N$ simple unitary Wilson loop matrix $W$, 
defined on a curve that does not self intersect, with $\tau$
denoting the dimensionless area in units of the 't Hooft gauge coupling,
has the following probability distribution:
\be
P_N (W,\tau)dW = \sum_R d_R \chi_r (W) e^{-\tau C_2 (R)} dW
\ee
The sum is over all irreducible representations $R$ with character $\chi_R (W)$ and
second order Casimir $C_2 (R)$. $dW$ is the Haar measure. Normalization conventions
are standard~\cite{three-d} and $\tau\ge 0$. We introduce the average characteristic polynomial 
\be
Q_N(z,\tau)=\langle \det (z - W)\rangle_{P_N(\tau)}
\ee
One can think about $Q_N(z,\tau)$ as the generating 
function for the 
$\langle \chi_R (W)\rangle$ with totally antisymmetric $R$.  
Simple manipulations~\cite{three-d} produce an integral representation:
\be
Q_N (z,\tau)=\sqrt{\frac{N\tau}{2\pi}} \int_{-\infty}^\infty du e^{-\frac{N}{2} \tau u^2} \left [z -e ^{-\tau(u+1/2)}\right ]^N
\ee
It is more convenient to study
\be
q_N(y,\tau) = (-1)^N e^{-\frac{Ny}{2}} e^{\frac{N\tau}{8}} Q_N (-e^y,\tau)
\label{qpol}
\ee
where, for the time being, $y$ is kept real. $q_N (y,t)$ is even in $y$ and this is
the main reason for extracting the exponential factor from $Q_N$. Changing the integration 
variable $u$ to $x=y+\tau (u+1/2)$ gives:
\be
q_N(y,\tau)= \sqrt{\frac{N}{2\pi\tau}} \int_{-\infty}^\infty dx e^{-\frac{N}{2\tau}  (y-x)^2} e^{N\log (2\cosh(x/2))}
\ee

\section{Main result.}

It is now a trivial matter to observe that
\be
\frac{\partial q_N}{\partial \tau} =\frac{1}{2N}\frac{\partial^2 q_N}{\partial y^2}
\label{primaryb}
\ee
with initial condition
\be
\lim_{\tau\to 0} [q_N (y,\tau)]= (2\cosh (y/2))^N
\ee
The behavior at $y\to\pm\infty$ prevents solving (\ref{primaryb}) by
Fourier decomposition and any associated general conclusions about 
boundedness as $\tau\to+\infty$. The initial condition is a consequence of
\be 
P_N(W,0)=\delta(W,{\bf 1})~~{\rm with}~~ \int dW \delta(W,W_0) f(W)=f(W_0)
\ee 
for any $W_0\in SU(N)$. This equation can be also directly derived from the polynomial
formula of $Q_N$, without going to the integral representation. 
This heat equation is related 
to Burgers' equation (for example, see~\cite{fritz}, 
problem 12(a), p. 214) by 
\be
\phi_N(y,\tau)=-\frac{1}{N}\frac{\partial \log q_N(y,\tau)}{\partial y}
\ee
Burgers' equation and the initial condition are
\be
\frac{\partial \phi_N}{\partial \tau} +\phi_N\frac{\partial \phi_N}{\partial y}=\frac{1}{2N}\frac{\partial^2 \phi_N}{\partial y^2},~~~ \phi_N (y,0)=-\frac{1}{2}\tanh\frac{y}{2}
\label{primarya}
\ee

At $N=\infty$, $N$ drops out of the equation giving the inviscid limit:
\be
\frac{\partial \phi}{\partial \tau} +\phi\frac{\partial \phi}{\partial y}=0
\ee
The initial condition is $N$ independent so we can drop the $N$ subscript on 
$\phi$ at $N=\infty$. So long as $\phi$ is uniquely defined, this is the
point-wise $N=\infty$ limit of $\phi_N$. 

The equation can be solved by the method of 
characteristics (for example, see ~\cite{fritz}, p.  16.) 
for an arbitrary initial condition
\be
\phi(y,0)=h(y)
\ee
The solution is given implicitly by
\be
\phi(y,\tau)=h(y-\tau\phi(y,\tau))
\ee
This equation is known to produce a shock at a time $\tau^* >0$ which is the
first time at which multiple solutions become available. $\tau^*$ is the smallest
positive value satisfying
\be
\tau^*=-\frac{1}{(dh/dy) (y^*)}~~{\rm with}~~(d^2 h/dy^2)(y^*)=0
\ee

We are interested only in solutions odd in $y$; hence, assuming $h(y)$ 
to be smooth near $y=0$ we expand:
\be
h(y)=a y + b y^3 + c y^5 +....
\label{initca}
\ee
This implies that $y^*=0$ and therefore
\be
\tau^*=-\frac{1}{a}
\ee
A shock will form if $a < 0$. In the case of $N=\infty$ 2D YM we
have
\be
h(y)=-\frac{1}{2}\tanh\frac{y}{2}=-y/4 +y^3/48-....
\ee
Therefore, the critical area corresponds to
\be
\tau^*=4, 
\ee
the well known critical value~\cite{duol,janik}. 

Universality can be invoked now in a sense that applies to the nonlinear equation
producing a generic shock~\cite{bess-four,caflisch}. This means taking
the simplest polynomial $h(y)$ capable of producing shocks:
\be
h(y)=ay+by^3
\label{initcb}
\ee
with $a<0,~b>0$. The $y$ location of the shock is at the origin, $y=y^*=0$.
Extending $h$ and $y$ to the complex plane provides a geometric view of this universality 
in terms of the structure of the evolving Riemann surface $y(\phi,\tau)$ 
parameterized by $\tau\ge 0$. One can also take $\tau$ into the complex plane. 

\section{Large $N$ = small viscosity.}

Making the viscosity nonzero is a singular perturbation which eliminates the
shock and has the same effect as making $N$ finite. Large $N$ universality
will hold in the vicinity of the critical area and corresponds to  
universal behavior in the vicinity of the would-be shock for small viscosities,
which is the simplest dissipative\footnote{
The shock can be regulated also
dispersively, in which case we could use a third derivative on the right hand side
of the inviscid Burgers' equation, producing the KdV equation.
If there were a symmetry 
restricting to a Hamiltonian partial differential equations, this might
have been the equation defining the universality class.} 
regularization of the shock. 

The important
new insight is that the large $N$ transition is equivalent to 
a movable singularity, determined by 
the initial condition, rather than by the evolution rule.\footnote{Something similar happens in the context of models consisting of one or several large
matrices, where Painlev{\'{e}} equations enter (see for example~\cite{makeenko} and
~\cite{ablowitz}).}
Thus, the simplest initial
condition producing a shock will also 
lead to a universal small viscosity smoothing
of the shock.

Running the derivation backwards, with the minimal initial condition
\be
h(y)=-y/4+y^3/48
\ee
produces an integral representation on which a double scaling limit can be taken
directly, exactly reproducing the limit used in matching to the
large $N$ transitions in higher dimensions than two in~\cite{ourjhep,three-d}.
The critical exponents $\mu=1/2,3/4$ 
associated with the scalings $N^\mu$ that need to
be taken~\cite{three-d} are identical to those found in defining the small viscosity limit~\cite{senouf}. The associated integral, studied in detail in~\cite{three-d} 
(see ~\cite{lat07} for a plot), 
is related to Pearcey's integral by a contour change, as indicated in~\cite{blaizot}.

The particular initial condition~(\ref{initcb}) has been analyzed in great detail in~\cite{senouf}. 

\section{Higher critical points.}

We have become accustomed to expect higher critical points, of reduced
degrees of stability, to accompany a basic large $N$ critical point. Looking at
(\ref{initca}) it seems plausible that setting $b=0$ and making $c>0$ would
produce a critical point of one degree of stability less. Obviously, if this
is true, a whole hierarchy will be generated, by initial conditions of the
form $ay+by^{2m-1}$ with integer $m\ge 2$. If one is 
not worried about the convergence of the associated universal integrals and one is
also willing to give up the $y\to -y$ parity symmetry, also higher 
critical points with half integer $m$
could be studied, at least as formal originators of asymptotic series. 

It would be intriguing if parent models existed with 
physical symmetries that selected one of these higher critical points. 
More work on this is left for the future.

\section{Product representation.}

It certainly is true that

\be
Q_N(z,\tau)=\langle \det (z - W)\rangle_{P_N(\tau)}=\prod_1^N (z-z_i(\tau))
\label{qpolb}
\ee

One may view the $z_i(\tau)$ as certain averages of the eigenvalues of $W$, but
not as usually defined:
\be
\det (z - W)=\prod_1^N (z-{\hat z}_i(W)),~~~~~{\hat z}^{av}_i (\tau) =
\langle {\hat z}_i (W) \rangle_{P_N (\tau )}
\ee
In~\cite{three-d} it was proved that $|z_i(\tau)|=1$
for all $i=1,...,N$ (see~\cite{lat07} for a plot); this indicates that 
the $z_i(\tau)$ are to be viewed as an approximations to the ${\hat z}^{av}_i(\tau)$. 
By applying large $N$ factorization $N$ times, 
one can argue, at least away from large $N$ critical points, 
that identically ordered ${\hat z}^{av}_i(\tau)$'s and $z_i(\tau)$'s
are equal to each other.  
I suspect that this stays true also in the double scaling limit.  
If this suspicion is validated, we shall obtain a new method 
to identify, using numerical
simulations, the location and  nature 
of the large $N$ transition in dimensions higher than two. 

It is therefore interesting to derive evolution equations for the
$z_i(\tau)$. After inserting the product 
(\ref{qpolb}) into the heat equation (\ref{primaryb}) and applying 
(\ref{qpol}), standard manipulations of the kind employed in the study of
Calogero systems produce
\be
\frac{{\dot z}_j}{z_j} =\frac{1}{2N}{\sum_{k}}' \frac{z_k+z_j}{z_k-z_j},~~~{\rm for}~~j=1,..,N
\label{eqmot}
\ee
Here ${\dot z}_j=dz_j (\tau )/d\tau$ and ${\sum_{k}}'$ means that the $k=j$ term 
is dropped from the sum, where the index $k$ runs from $1$ to $N$. 
This equation is form invariant under $z_j \to 1/z_j$ and $z_j\to z_j^*$, as expected from the structure of the polynomial. In addition, 
again as expected, the product of all zeros is constant in $\tau$. Moreover, 
the equations of motion (\ref{eqmot}) imply $d|z_j(\tau)|^2/d\tau=0, j=1,..,N$. 

The initial condition is
$z_j(0)=1,~j=1,..,N$ and is degenerate.
However, at any $\tau >0$ the degeneracy is 
lifted; for example, at infinite $\tau$, we have $z_j(\infty)=e^{2\pi i (j+1/2-N/2)/N}$.

The map $z=-e^y$ creates an infinite number of copies of the zeros $z_j$,  
which are all on the imaginary axis. We choose one specific $y_j$ for
each $z_j, j=1,..,N$. The equation of motion for the $y_j$'s is:
\be
{\dot y}_j =\frac{1}{2N}{\sum_{k}}' \coth \frac{y_k-y_j}{2}=\frac{1}{N} {\sum_{k}}'
\sum_{n\in Z} \frac{1}{y_k - y_j + 2n\pi i }
\ee
The universal description changes the equation obeyed by the $y_j$'s. However,
as pointed out in~\cite{senouf} on the basis of an old theorem~\cite{polya}, the
$y_j(\tau)$ still stay on the imaginary axis for all $\tau$. In the universal case
periodicity under $y_j(\tau)\to y_j(\tau)+2m_j\pi i,~m_j\in Z$ is lost, since the initial condition on the $y_i$'s no longer is periodic. Thus, one needs to use the $y_j$ variables to make
the connection between the exact equations of motion and the universal ones. I leave a more
detailed study of the universal limit of the eigenvalue motion to the future.

\section{Large $\tau$ behavior.}

The regularization of the shock provides a smooth connection between small
and large loops. In two dimensions Burgers' equation provides an exact
renormalization group type of equation allowing the evaluation
of $\phi_N(y,\tau)$ when $\tau\to\infty$, given $\phi_N(y,\tau)$ in the 
limit $\tau\to 0$. The approach to the limit $\tau\to\infty$ gives the
dimensionless string tension associated with the dimensionless area $\tau$.
Here we only show how the correct $\phi_N(y,\tau=\infty)$ is obtained. 
It is clear that $Q_N(z,\tau=\infty)=z^N+(-1)^N$. This simply says that at infinite 
$\tau$ all $\langle W^m\rangle$ terms, for any $m>0$, can be replaced by zero. 

Using~(\ref{qpol}), we conclude that the large $\tau$ behavior is given by:
\be
\lim_{\tau\to\infty} \left ( e^{-\frac{N\tau}{8}} q_N(y,\tau)\right )=
2\cosh\frac{Ny}{2}
\ee
 
We now wish to recover the ensuing $\phi_N(y,\tau=\infty)$ from Burgers'
equation. The route is again in reverse of our derivation: First go to the
heat equation, then get the integral representation in order to incorporate
the initial condition. Finally, in order to get the asymptotic behavior
for large $\tau$, change variables in the integral representation, arriving at:
\begin{eqnarray}&
\frac{1}{N}\partial_y \log q_N (y,\tau)=\cr & 
\frac{\int du e^{-\frac{Nu^2}{2} }\sinh((u\sqrt{\tau}+y)/2) (2\cosh((u\sqrt{\tau}+y)/2))^{N-1}}{\int du e^{-\frac{Nu^2}{2}} (2\cosh((u\sqrt{\tau}+y)/2))^N }
\end{eqnarray}
For large $\tau$, one of the two exponents making up each hyperbolic function
dominates, depending on the sign of $u$:
\[
\lim_{\tau\to\infty}\left (\frac{1}{N}\partial_y \log q_N (y,\tau)\right )=
\]
\be
\frac{1}{2} \lim_{\tau\to\infty}\left ( \frac{\int du e^{-\frac{Nu^2}{2} }\varepsilon(u)
e^{N[\varepsilon(u)(u\sqrt{\tau}+y)/2]}}{\int du e^{-\frac{Nu^2}{2}}
e^{N[\varepsilon(u)(u\sqrt{\tau}+y)/2]}}\right )
\ee
Here, $\varepsilon(u)$ is the sign function. The above equation implies that
\be
\phi_N(y,\infty)=\lim_{\tau\to\infty}\left (-\frac{1}{N}\partial_y \log q_N (y,\tau)\right )
=-\frac{1}{2}\tanh\frac{Ny}{2}
\ee
This is the expected result.

At infinite $N$, the hyperbolic tangent becomes a sign function.
In an electrostatic picture it is obvious that the above result holds if
the poles of $\phi(y,\tau)$ are uniformly spaced and dense on the circle $|e^y|=1$: 
Viewing the poles as charges, the jump $\varepsilon(y)$ comes from crossing
the line charge at $z=-1$ as $y$ goes through zero along the real axis~\cite{three-d}. That the solution has this limiting behavior is essential
for confinement, which would be indicated by the leading correction to the
above result being exponentially small in $\tau$. 

Note that $\tau$ was taken to infinity at finite $N$; the final
result admits a subsequent infinite $N$ limit. Had we
taken $N\to\infty$ first, we could have interpreted the shock, appearing first at $\tau=4$,
as a jump between two extremal solutions of the implicit equation defining the solution for
$\tau <4$. With the wrong initial conditions this 
jump might not grow to the full size required
for consistency with confinement; thus, the transition in itself is insufficient to
guarantee confinement. If we want to add the input that there is confinement we need to
put a constraint on the initial condition. 

Regarding 
~\cite{blaizot}, following ~\cite{frisch}, I opt not to
address here the question how Burgers' equation relates to turbulence. As a start, I refer the reader to ~\cite{burgers}.
In general, one would hope that the analogy to the 
three dimensional incompressible Navier-Stokes equation does not
hold too literally. Large $N$ would map to large Reynolds numbers, while small $N$ 
to small Reynolds numbers; however, I am
hoping that matters simplify at large $N$ -- if they do not, 
one would be better off concentrating on $N=3$.

Again, I leave details for further work.

\section{Discussion.}

The primary objective of this paper was the derivation of~(\ref{primarya}) as an
exact equation holding in two dimensional Yang Mills theory with gauge group 
$SU(N)$ defined on the infinite Euclidean plane. 
A surprising simplicity in the area dependence
of the average characteristic polynomial 
of simple Wilson loops was found. Nevertheless, the essential 
feature of the existence of a large $N$ phase transition is captured by this
observable. In this respect the average characteristic polynomial of the Wilson loop is superior to traces of the Wilson loop in some fixed representation. 
As explained in~\cite{three-d} this observable has other advantages, in dimensions
three and four. 

The simple and exact finite $N$ relation to Burgers' equation presented above seems
to provide opportunity for progress in different directions, as emphasized 
in the course of this paper. The secondary objective of the paper was to present
enough observations to convince the reader that there are many interesting issues
left to explore. Last, but not least, the insights of Blaizot 
and Nowak~\cite{blaizot} deserve further study. 

The shock at $\tau=4$ is reminiscent of the 
possibility that instantons at infinite $N$ might herald, as $\tau\to 4^-$, a jump 
in certain particularly sensitive quantities in 4D YM ~\cite{inst}.

It should also be mentioned that workers in lattice field theory~\cite{thies} have
shown numerically that in four dimensions the trace $2\cos\theta$ 
of a Wilson loop for $SU(2)$ seems to evolve with the area as if $\theta$ were
diffusing on the $SU(2)$ group manifold where the eigenvalues of $W$ are $e^{\pm i\theta}$.
For $N=2$ there is no essential distinction between the characteristic polynomial 
and any other gauge invariant observable related to the matrix $W$.

\section{Added note.}

Blaizot and Nowak~\cite{newbn} have independently 
identified the viscosity as $\frac{1}{2N}$.

\vskip 4mm

\subsection*{Acknowledgments.}

I acknowledge partial support by the DOE under grant
number DE-FG02-01ER41165 at Rutgers University and by the SAS of Rutgers University.
I note with regret that my research has for a long time been 
deliberately obstructed by my high energy colleagues at Rutgers.  
I gratefully acknowledge an email from the authors of ~\cite{blaizot} 
on Feb. 27, 2008, drawing my attention 
to their work and a further email commenting on the
present manuscript and identifying seminars where 
they presented their work in progress. 
An ongoing collaboration on related topics
with R. Narayanan, as well as comments on this manuscript, 
are also gratefully acknowledged. I also wish to thank J. Feinberg for
some comments on the manuscript.

\end{document}